\documentclass{article}
\usepackage{epsfig}

\date{\today}

\begin{document}
\begin{center}
{\Large\bf Gravitating Yang-Mills dyon vortices \\in 4+1 spacetime dimensions}
\vspace{0.5cm}
\\
{\bf Yves Brihaye$^{1}$ }
{\bf and Eugen Radu$^{2}$ }

\vspace*{0.2cm}
{\it $^1$Physique-Math\'ematique, Universite de
Mons-Hainaut, Mons, Belgium}

{\it $^2$ Department of
Mathematical Physics, National University of Ireland Maynooth, Maynooth, Ireland}
\vspace{0.5cm}
\end{center}
\begin{abstract}
We consider  vortex-type solutions in $d=5$ dimensions 
of the Einstein gravity coupled
to a nonabelian SU(2) field posessing a nonzero electric part.
After the dimensional reduction, this corresponds to a $d=4$
Einstein-Yang-Mills-Higgs-U(1)-dilaton model.
A general axially symmetric ansatz is presented, and
the properties of the spherically symmetric solutions  are analysed.
\end{abstract}

\section{Introduction}
Following the
discovery by Bartnik and McKinnon of
particle-like solutions of the four-dimensional
Einstein-Yang-Mills (EYM) equations
\cite{Bartnik:1988am}, there has been much interest
in classical solutions
of Einstein gravity  with nonabelian matter sources.

A recent progress in this area was due
to investigation of nontrivial configurations
in higher dimensional spacetimes,
particle-like and black hole solutions of  EYM theories
being studied in \cite{Brihaye:2002hr, Brihaye:2002jg}.
These configurations are sphericaly symmetric in the $d-$dimensional spacetime
and are sustained by the higher order terms in the Yang-Mills (YM) hierarchy.
Without these terms, only vortex-type solutions,
with a number of codimensions, are possible to exist.

The situation is exemplified by
the best understood EYM-SU(2) model in
$4+1$ dimensions.
Without gravity, the pure YM theory in $d=5$ Minkowski spacetime
admits topologically stable, particle-like  and
vortex-type solutions obtained by uplifting the $d=4$ YM instantons and
$d=3$ YM-Higgs monopoles.
However, as found in \cite{Volkov:2001tb},
the particle spectrum become completely destroyed 
by gravity,
while the vortices admit very non-trivial gravitating generalizations.
Assuming the metric and matter fields to be independent on the extra coordinate $x^{5}$,
 then the
($4+1$)-dimensional static EYM system reduces effectively
to a ($3+1$)-dimensional Einstein-Yang-Mills-Higgs-dilaton (EYMHD) system.
The solutions of this system exist for a finite range of values of the
gravitational coupling constant,
and in the strong gravity limit they become
gravitationally closed.
These gravitating vortices comprise an infinite family 
including the fundamental solution and its excitations.

These ideas were taken further
to $4+n$ dimensions \cite{Brihaye:2004kd}, where $n$ Higgs triplets
and $n$ dilatons appear.
For two or more extra dimensions,
a number of constraints
result from the off-diagonal terms of the energy-momentum tensor, 
which implies a more complicated metric ansatz.

Here we remark that all these studies are resticted to a purely magnetic YM  connection.
In the four dimensional case, the electric YM potential
necessarily vanishes for
static asymptotically flat regular EYM solutions 
\cite{Bizon:1992pi,Ershov:1991nv}.\footnote{
However, gravitating dyon solutions  exist in the presence of a Higgs field in the
adjoint representation \cite{Brihaye:1998cm}.}

Here we present numerical arguments that
this result do not generalize to higher dimensions.
We show that the known vortex-type solutions admit dyonic generalizations,
existing up to some maximal value of the gravitational coupling constant.
Beside the fundamental gravitating dyon solutions, we also construct
excited configurations.
To simplify the general picture, we consider the case of 4+1 dimensions only,
although similar results exist for any $d>4$.

\section{The model}
\subsection{Action principle  }
The five dimensional EYM-SU(2) coupled system is described by the action
\begin{equation}
\label{action5}
I_5=\int d^{5}x\sqrt{-g }\Big(\frac{R }{16\pi G}
-\frac{1}{2g^2}Tr\{F_{MN }F^{MN} \}\Big),
\end{equation}
(throughout this letter, the indices $\{M,N,...\}$ will
indicate the five dimensional
coordinates and $\{\mu,\nu,...\}$ will indicate the
coordinates in the four dimensional
physical spacetime; also, the length of the extra dimension $x^5$ is taken to be one).

Here $G$ is the gravitational constant, $R$ is the Ricci scalar associated with the
spacetime metric $g_{MN}$
and
$F_{MN}=\frac{1}{2} \tau^aF_{MN}^{(a)}$ is the gauge field strength tensor defined as
\begin{equation}
F_{MN} =
\partial_M A_N -\partial_N A_M - i   \left[A_M , A_M \right],
\label{fmn}
\end{equation}
where the gauge field is
$A_{M} = \frac{1}{2} \tau^a A_M^{(a)},$
$\tau^a$ being the Pauli matrices and $g$ the gauge coupling constant.

Variation of the action (\ref{action5})
 with respect to the metric $g^{MN}$ leads to the Einstein equations
\begin{equation}
\label{einstein-eqs}
R_{MN}-\frac{1}{2}g_{MN}R   = 8\pi G  T_{MN},
\end{equation}
where the YM stress-energy tensor is
\begin{eqnarray}
T_{MN} = 2{\rm Tr}
    ( F_{MP} F_{NQ} g^{PQ}
   -\frac{1}{4} g_{MN} F_{PQ} F^{PQ}).
\end{eqnarray}
Variation with respect to the gauge field $A_\mu$
leads to the YM equations
\begin{equation}
\label{YM-eqs}
\nabla_{M}F^{MN}-i[A_{M},F^{MN}]=0.
\end{equation}
In \cite{Volkov:2001tb} is has been proven  that
 there are no purely magnetic ($A_t=0$)
finite-energy, particle-like solutions of the above equations with a
SO(4) spacetime symmetry group.
Also, as explicitly shown in the Appendix 1 of Ref. \cite{Okuyama:2002mh},
a spherically symmetric nontrivial YM-SU(2) ansatz with a
nonzero electric potential
is necessarily time dependent.
To avoid these  results, we are forced to modify the original action
by  consider
higher order terms in the YM hierarchy \cite{Brihaye:2002hr}, \cite{Brihaye:2002jg}.

\subsection{The ansatz}
However,  the action principle (\ref{action5})
presents less symmetric solutions with interesting properties.
In what follows we will consider vortex-type configurations,
assuming that both the matter functions and
the metric functions are
independent on the extra-coordinate $x^5$.
 Without any loss of generality, we consider a five-dimensional metric parametrization
\begin{eqnarray}
\label{metrica}
ds^2 = e^{- 2\phi/\sqrt{3}}\gamma_{\mu \nu}dx^{\mu}dx^{\nu}
 + e^{ 4\phi/\sqrt{3}}(dx^5 + 2W_{\mu}dx^{\mu})^2.
\end{eqnarray}

With this assumption, the considered theory admits an interesting Kaluza-Klein (KK) picture.
While the  KK reduction of the Einstein term in  (\ref{action5}) is standard,
to reduce the YM action term,
is convinient to take an  SU(2) ansatz
\begin{eqnarray}
\label{SU2}
A={\cal A}_{\mu}dx^{\mu}+g\Phi (dx^5+2 W_\mu dx^\mu),
\end{eqnarray}
where $W_\mu$ is a U(1) potential,
${\cal A}_{\mu}$ is a purely four-dimensional gauge field potential,
while  $\Phi$ corresponds after the dimensional reduction to a triplet Higgs field.

This leads to the four dimensional action principle
\begin{eqnarray}
\label{action4}
I_4=\int d^{4}x\sqrt{-\gamma }\Big[
\frac{1}{4\pi G}\big(
\frac{\mathcal{R} }{4}
-\frac{1}{2}\nabla_{\mu}\phi \nabla^{\mu}\phi
-e^{2\sqrt{3}\phi}\frac{1}{4}G_{\mu \nu }G^{\mu \nu } \big)
-e^{2\phi/\sqrt{3}}\frac{1}{2g^2}Tr\{
{\cal F}_{\mu \nu }{\cal F}^{\mu \nu }\}
\\
\nonumber
-e^{-4\phi/\sqrt{3}}Tr\{ D_{\mu}\Phi D^{\mu}\Phi\}
- 2 e^{2\phi/\sqrt{3}}\frac{1}{g}G_{\mu \nu} Tr\{\Phi {\cal F}^{\mu \nu} \}
-2e^{2\phi/\sqrt{3}} G_{\mu\nu}G^{\mu\nu}Tr\{  \Phi^2 \}
\Big],
\end{eqnarray}
where $\mathcal{R}$ is the Ricci scalar for the metric $\gamma_{\mu \nu}$,
while
 ${\cal F}_{\mu \nu }=
\partial_{\mu}{\cal A}_{\nu}
-\partial_{\nu}{\cal A}_{\mu}-i [{\cal A}_{\mu},{\cal A}_{\nu}  ]$
and
 $G_{\mu \nu}=\partial_{\mu}W_{\nu}-\partial_{\nu}W_{\mu}$
are the SU(2) and U(1) field strength tensors defined in $d=4$.

Here we consider five dimensional configurations possessing two more Killing vectors
apart from $\partial/\partial x^5$,
$\xi_1=\partial/\partial \varphi$,
corresponding to an axially
symmetry of the four dimensional metric sector
(where the azimuth angle $\varphi$
 range from $0$ to $2 \pi$),
and $\xi_2=\partial/\partial t$,
with $t$ the time coordinate.

The five dimensional  YM ansatz in this case
 is a straightforward generalization of the axially symmetric
 $d=4$ ansatz
obtained in the pioneering papers by Manton \cite{Manton:1977ht}
and Rebbi and Rossi \cite{Rebbi:1980yi}.
For the time and extra-direction
translational symmetry, we choose a gauge such that
$\partial A/\partial t=\partial A/\partial x^5=0$.
However, a $\varphi-$rotation around the $z-$axis
can be compensated by a gauge rotation
\begin{eqnarray}
\label{Psi}
{\mathcal{L}}_{\varphi} A_{N}=D_{N}\Psi,
\end{eqnarray}
with $\Psi$ being a Lie-algebra valued gauge function.
This introduces an winding number $n$ in the ansatz (which is a constant of motion
and is restricted to be an integer)
and implies the existence of a potential $W$ with
\begin{eqnarray}
\label{relations}
F_{N \varphi} =& D_{N}W,
\end{eqnarray}
where $W=A_{\varphi}-\Psi$.

The most general axially symmetric 5D Yang-Mills ansatz contains 15 functions: 12 magnetic
and 3 electric potentials and can be easily obtained in cylindrical coordinates
$x^{N}=(\rho,\varphi,z$) (with $\rho=r\sin \theta,~z=r\cos\theta,$
and $r$, $\theta$ and $\varphi$ being the usual spherical coordinates)
\begin{eqnarray}
\label{A-gen-cil}
A_{N}=\frac{1}{2}A_{N}^{(\rho)}(\rho,z)\tau_{\rho}^n
        +\frac{1}{2}A_{N}^{(\varphi)}(\rho,z)\tau_{\varphi}^n
        +\frac{1}{2}A_{N}^{(z)}(\rho,z)\tau_{z},
\end{eqnarray}
where the only $\varphi$-dependent terms are the SU(2) matrices
(composed of the standard $(\tau_x,~\tau_y,~\tau_z)$ Pauli matrices)
\begin{eqnarray}
\label{u-cil}
\tau_{\rho}^n=~~\cos n\varphi~\tau_x+\sin n\varphi~\tau_y,
~~
\tau_{\varphi}^n=-\sin n\varphi~\tau_x+\cos n\varphi~\tau_y,
\end{eqnarray}
(we can reduce this general ansatz by imposing some extra symmetries on the fields).

Transforming to
spherical coordinates, it is convenient to introduce,
without any loss of generality, a new SU(2) basis
$(\tau_{r}^n,\tau_{\theta}^n,\tau_{\varphi}^n)$,
with
\begin{eqnarray} 
\label{u-sph}
\tau_{r}^n=\sin \theta~\tau_{\rho}^n+\cos \theta~\tau_z,
~~
\tau_{\theta}^n=\cos \theta~\tau_{\rho}^n-\sin \theta~\tau_z,
\end{eqnarray}
which yields the general expression
\begin{eqnarray}
\label{A-gen-sph}
A_{N}=\frac{1}{2}A_{N}^{r}(r,\theta)\tau_{r}^n
        +\frac{1}{2}A_{N}^{\theta}(r,\theta)\tau_{\theta}^n
        +\frac{1}{2}A_{N}^{\varphi}(r,\theta)\tau_{\varphi}^n.
\end{eqnarray}
For this parametrization
%
$2\Psi=n \tau_z =n \cos \theta  \tau_{r}^n
- n \sin \theta  \tau_{\theta}^n.$
The gauge invariant quantities expressed in terms of these functions
will be independent on the
angle $\varphi$.

 The assumed symmetries together with
the YM equations implies the following relations
for some extradiagonal components of the
energy-momentum tensor
\begin{eqnarray}
\label{t1}
T_t^5&=&2Tr\{ \frac{1 }{\sqrt{-g}}\partial_{N} (\sqrt{-g}  A_t F^{N5} )\},
~~
T_5^t=2Tr\{ \frac{1 }{\sqrt{-g}}\partial_{N} (\sqrt{-g}  A_5 F^{Nt} )\},
\\
\nonumber
T_{\varphi}^{t}&=&2Tr \{\frac{1}{\sqrt{-g}}\partial_{N}(\sqrt{-g}W F^{N t} )\},
~~
T_{t}^{\varphi}=2Tr\{ \frac{1 }{\sqrt{-g}}\partial_{N} (\sqrt{-g}  A_t F^{N \varphi} )\}.
\end{eqnarray}
Also, one  can prove the results
\begin{eqnarray}
\label{rels}
E_{e}&=&2 Tr\{\int_V d^3x   \sqrt{-g} F_{M t}F^{M t} \}=
2 Tr\{\oint_{\infty}  d S_{\mu} \sqrt{-g} A_t F^{\mu t} \},
\\
E_{h}&=&2 Tr\{\int_V d^3x  \sqrt{-g} F_{M 5}F^{M 5} \}=
2 Tr\{\oint_{\infty} d S_{\mu} \sqrt{-g} A_5 F^{\mu 5}  \},
\\
\label{j5}
J &=&2 Tr\{ \int_V d^{3}x\sqrt{-g}
F_{M \varphi}F^{M t} \}
=2Tr\{\oint_{\infty}dS_{\mu}~\sqrt{-g}WF^{\mu t} \},
\end{eqnarray}
(where the volume integral is taken over the three dimensional physical
space)
which implies that, for nontrivial dyonic particle-like solutions,
the magnitude of the gauge potentials $A_t,~A_5$
should be nonzero at infinity.

In this letter,  we will restrict to a spherically symmetric ansatz
for the four dimensional metric
\begin{eqnarray}
\label{metrica4}
\gamma_{\mu \nu}dx^\mu dx ^\nu= \frac{dr^2}{N(r)}
+ r^2(d \theta^2 + \sin^2 \theta  d \phi^2 )- N(r) \sigma^2(r) dt^2
\end{eqnarray}
with
\begin{eqnarray}
\nonumber
N(r)=1-\frac{2m(r)}{r}.
\end{eqnarray}
This implies the existence of two more Killing
vectors apart from 
$\partial/\partial x^5,~\partial/\partial t,~\partial/\partial \varphi$,
originating in the SO(3) symmetry of the above line element.

The expression of the five-dimensional
SU(2) connection
is obtained by using the standard rule
for calculating the gauge potentials for any spacetime group
\cite{Forgacs:1980zs,Bergmann,Basler:yr}.
Taking into account
the symmetries of the metric form (\ref{metrica}),
a straightforward computation
leads to the YM ansatz
\begin{eqnarray}
\label{A5}
A_r &=&0,~~A_{\theta}=(1-K(r))\tau_{\varphi}^{1 },
~~A_{\varphi} =-(1-K(r))\sin \theta\tau_{\theta}^{1 },
\\
\nonumber
A_t &=&\Big(u(r)+2J(r)H(r)\Big)\tau_r^1,~~A_5 =H(r)\tau_r^{1 },
\end{eqnarray}
where  $(\tau _r^1,~\tau _\theta^1,~\tau _\varphi^1)$
are found by taking $n=1$ in the expressions (\ref{u-cil}), (\ref{u-sph}).

We remark that
$A_t,~A_5$ are oriented along the same direction in the isospace.
Therefore, the energy-momentum tensor of nontrivial YM configurations presents
also $T_{t}^5,~T_5^t$ components, while $T_\varphi^t=T^\varphi_t=0$.
This implies the existence, in the five dimensional metric ansatz (\ref{metrica}),
of one extradiagonal component
\begin{eqnarray}
W_{\mu}=J(r)\delta_{\mu}^t,
\end{eqnarray}
corresponding,
in a four dimensional picture,
to an U(1) electric potential.
Thus, any five dimensional SU(2) dyon configuration necessarily has 
a nonzero $J(r)$, which cannot be gauged away.
A vanishing four dimensional electric potential implies a purely magnetic
SU(2) configuration in $d=5$.

In the same approach, $K(r),~u(r)$ are
$d=4$ 
 magnetic and electric SU(2)-gauge potentials, $\phi(r)$ is a dilaton while $H(r)$ is the
Higgs field.
The configurations discussed in this paper extremize also the
action principle (\ref{action4}) and can be viewed
as regular solutions of the four dimensional
theory. The existence  of a nonvanishing Abelian electric
field in $d=4$ without a singular source is not a surprise, given the specific coupling
of the U(1) potentials to SU(2) and Higgs fields in the action principle (\ref{action4}).
A similar property has been noticed in other models with a Maxwell field
interacting with some matter fields (see e.g. \cite{Jetzer:1989av}).

\subsection{The equations of motion and boundary conditions}
Dimensionless quantities are obtained by using the following rescaling
\begin{eqnarray}
\label{resc}
r \to r/(g H_0), ~~H(r) \to g H_0H, ~~u(r) \to g H_0 u,~~m(r) \to m/(g H_0),
\end{eqnarray}
where $H_0$ is the asymptotic value of $H(r)$
and we define also the
coupling constant $\alpha=H_0 \sqrt{4 \pi G}$
(there also
nontrivial solutions
with $H_0=0$; however, that implies a purely magnetic ansatz $u(r)=0$).

With these conventions, the field equations 
 for the four metric variables
$(m,~\sigma,~\phi,~J)$ and 
the three matter functions $(K,~H,~u)$ take the form
\begin{eqnarray}
\label{e1}
\nonumber
m'&=&e^{2\sqrt{3}\phi}\frac{r^2 J'^2}{2\sigma^2}+
\frac{1}{2}r^2N\phi'^2+
\alpha^2\Big(
e^{ 2\phi /\sqrt{3}}T_1+
e^{- 4\phi/\sqrt{3}}T_2
+\frac{e^{ 2\phi/\sqrt{3}}}{\sigma^2}
T_3 \Big),
\\
\nonumber
\label{e2}
\frac{\sigma'}{\sigma} &=&   r  \phi'^2
+\frac{2\alpha^2}{r} \Big(
 e^{ 2\phi/\sqrt{3}} K'^2 +
  e^{- 4\phi/\sqrt{3}} \frac{r^2H'^2}{2}
+\frac{e^{ 2\phi/\sqrt{3}}2}{ \sigma^2 N^2}
K^2u^2
\Big)
\\
\label{e3}
\nonumber
(e^{ 2\phi/\sqrt{3}}\sigma NK')'&=&
e^{ 2\phi/\sqrt{3}}\sigma \frac{K(K^2-1)}{r^2}
+e^{ -4\phi/\sqrt{3}}\sigma KH^2
-\frac{e^{ 2\phi/\sqrt{3}}Ku^2}{\sigma N},
\\
\label{e5}
\nonumber
(r^2\sigma N\phi')'&=&-\sqrt{3}
e^{ 2\sqrt{3}\phi }\frac{r^2 J'^2}{\sigma}
+\frac{2\sigma\alpha^2}{\sqrt{3}}
\Big(
e^{ 2\phi/\sqrt{3}}  T_1
-2e^{ -4\phi/\sqrt{3}}  T_2
-e^{ 2\phi/\sqrt{3}}\frac{T_3}{\sigma^2}
\Big),
\\
\label{e4}
\Big(e^{ 2\sqrt{3}\phi }\frac{r^2 }{\sigma}(u'+2J'H)
\Big)'&=&
2e^{ 2\phi/\sqrt{3}}\frac{K^2u}{\sigma N},
\\
\nonumber
\label{e6}
0&=&\Big (e^{ 2\sqrt{3}\phi }\frac{r^2J'}{\sigma}
+2\alpha^2e^{ 2\phi/\sqrt{3}}\frac{r^2H}{\sigma}
(u'+2J'H) \Big)'
\\
\label{e7}
\nonumber
(e^{ -4\phi/\sqrt{3}}\sigma Nr^2H')'&=&
2e^{ -4\phi/\sqrt{3}}\sigma K^2H+2e^{ 2\phi/\sqrt{3}}
\frac{r^2J'}{\sigma}(u'+2J'H)^2,
\end{eqnarray}
where
\begin{eqnarray}
\nonumber
T_1=NK'^2+\frac{(K^2-1)^2}{2r^2},~~
T_2=\frac{1}{2}Nr^2H'^2 +K^2H^2,~~
T_3=\frac{r^2}{2}(u'+2J'H)^2 +\frac{K^2u^2}{N}.
\end{eqnarray}

Here we notice the following expressions for the extradiagonal
components of the Einstein  tensor
\begin{eqnarray}
\label{Ricci}
E_t^5=\frac{1}{\sqrt{-g}}\frac{d}{dr}(\Gamma_{t M}^rg^{M 5}\sqrt{-g}),
~~E_5^t=\frac{1}{\sqrt{-g}}\frac{d}{dr}(\Gamma_{5M}^rg^{M t}\sqrt{-g}),
\end{eqnarray}
which, together with (\ref{t1})
imply the existence of two first first integrals of
the field equations.
For regular configurations,
we find  the simple relations
\begin{eqnarray}
\nonumber
u'&=&-\frac{J'}{2\alpha^2 H}(e^{4\phi/\sqrt{3}}+4\alpha^2H^2),
\\
\label{rels1}
e^{4\phi/\sqrt{3}}\Big(N'J\sigma&+&
N(2J\sigma'-\sigma(J+2\sqrt{3}J\phi'))\Big)
-4e^{10\phi/\sqrt{3}}J'J^2
\\
\nonumber
&=&2\alpha^2
\Big(N\sigma^2 H'(u+2JH)+
2e^{2\phi/\sqrt{3}}J(u'+2JH')(u+2JH)
\Big).
\end{eqnarray}
An exact nonabelian dyon solution of the EYM equations can be
constructed, it reads
\begin{eqnarray}
\label{exact-sol1}
ds^2=\big((\beta^2+1)e^{c_3 x}-\beta^2e^{c_2 x}\big)
\Big(dx^5+2J(x)dt\Big)^2
+dx^2+c_1 (d \theta^2+\sin ^2 \theta d \varphi^2)
-\frac{e^{(c_2 +c_3)x}}{(\beta^2+1)e^{c_3 x}-\beta^2 e^{c_2 x}} dt^2,
\\
\nonumber
J(x)=\frac{1}{2}\frac{(e^{c_3 x}-e^{c_2 x}) \beta \sqrt{\beta^2+1}}
{(\beta^2+1)e^{c_3 x}-\beta^2e^{c_2 x}},~~K(x)=\sqrt{q},~~
H(x)=\sqrt{h}\sqrt{\beta^2+1}e^{c_3 x/2},~~u(x)=\beta\sqrt{h} e^{c_3 x/2}
-2 J(x)H(x),
\end{eqnarray}
where $\beta$ is an arbitrary real parameter and
%
%
\begin{eqnarray}
\label{exact-sol2}
\nonumber
c_1&=&\frac{1-q}{h}\simeq 1.12637 \alpha^2,~~
c_2=\frac{8 h q}{c_3(q-1)}\simeq \frac{1.26557}{\alpha},~~
c_3=2\sqrt{\frac{h\big(-1+h\alpha^2(1+q)\big)}{(q-1)(h\alpha^2-1)}}
\simeq \frac{0.482496}{\alpha},
\\
h&=&\frac{3+(3-2q)q+\sqrt{1+q\big(-6+q(-51+4(q-11)q)\big)}}{4(1+q)\alpha^2}
\simeq \frac{0.811482}{\alpha^2},
\\
\nonumber
q&=&\frac{1}{12}\Big(-7-\frac{83}{(1259+18\sqrt{6657})^{1/3}}
+(1259+18\sqrt{6657})^{1/3}\Big)\simeq 0.08597.
\end{eqnarray}
The coordinate $x$ used here is related to the radial coordinate $r$
defined in (\ref{metrica4}) by
$r^4 = \big((\beta^2+1)e^{c_3 x}-\beta^2e^{c_2 x}\big)$.
This generalizes  the "warped" $AdS_3\times S^2$ solution
found in \cite{Volkov:2001tb}
 which is recovered for $\beta=0$.
 
However, in this paper we will study globally regular, 
asymptotically flat solutions of the
field equations (\ref{e4}).
That implies
the following set of boundary conditions
\begin{eqnarray}
\label{c1}
K(0)=1,~~H(0)=0,~~\partial_r|_{r=0}\phi=0,~N(0)=1,~~\partial_r|_{r=0}J=0,
~~u(0)=0,~~\partial_r|_{r=0}\sigma=0,
\\
\label{c2}
K(\infty)=1,~~H(\infty)=1,~~\phi(\infty)=0,
~N(\infty)=1,~~ J(\infty)=0,
~~u(\infty)=\gamma,~~ \sigma(\infty)=1.
\end{eqnarray}
The asymptotic form of the solutions can be systematically
constructed in both regions, near the origin
and for large values of $r$.
The field equations implies
the following behavior as $r \to 0$ in terms of five parameters
$(b,j_0,u_1,h_1,\phi_0)$
\begin{eqnarray}
\nonumber
K(r)&=&1-br^2+O(r^4),~~
u(r)=u_1r+O(r^3),~~
H(r)=h_1r+O(r^3),
\\
\nonumber
N(r)&=&1-\frac{\alpha^2e^{ -4\phi_0/\sqrt{3}}}{\sigma_0^2}
\Big(h_1^2\sigma_0^2+e^{2\sqrt{3}\phi_0}(4b^2\sigma_0^2+u_1^2)\Big)r^2+O(r^3),
\\
J(r)&=&j_0-e^{ -4\phi_0/\sqrt{3}}\alpha^2 h_1u_1r^2+O(r^4),
\\
\nonumber
\sigma(r)&=&\sigma_0+\frac{e^{ -4\phi_0/\sqrt{3}}\alpha^2}{2\sigma_0}
\Big(h_1^2\sigma_0^2+2e^{2\sqrt{3}\phi_0}(4b^2\sigma_0^2+u_1^2)\Big)r^2+O(r^3),
\\
\nonumber
\phi(r)&=&\phi_0+\frac{e^{ -4\phi_0/\sqrt{3}}\alpha^2}{2\sqrt{3}\sigma_0^2}
\Big(-2h_1^2\sigma_0^2+e^{2\sqrt{3}\phi_0}(4b^2\sigma_0^2-u_1^2)\Big)r^2+O(r^3),
\end{eqnarray}
while a similar analysis as  $r \to \infty$ gives
\begin{eqnarray}
\nonumber
K(r)&\sim& e^{-\sqrt{1-\gamma^2} r},~~
J(r)\sim \frac{2Q\alpha^2\gamma}{r},~~
H(r)\sim1-\frac{Q}{r},~~
u(r)\sim \gamma(1-\frac{(1+4\alpha^2)Q}{r}),~~
\\
\phi(r)&\sim&\frac{\phi_1}{r},~~
N(r)\sim1-\frac{2M}{r},~~
\sigma(r)\sim 1-\frac{\phi_1^2+\alpha^2 Q^2}{2r},
\end{eqnarray}
where $\gamma,~Q,~M,\sigma_1,\phi_1$ are arbitrary parameters.
The parameter $M$ corresponds to
the ADM mass of the four dimensional solutions.
The nonabelian electric charge of the solutions is defined as $Q_e=(1+4\alpha^2)\gamma Q$,
while the $U(1)$ electric charge is just $2\alpha^2Q_e$.

For $u(\infty) > H(\infty)$
the asymptotic behaviour of $K(r)$  becomes oscillatory.
Therefore, the value of the electric potential at infinity
is restricted to be smaller than the asymptotic value of $A_5$, 
$i.e.$ $\gamma<1$ \footnote{A similar behavior has been noticed for
dyons in a
four dimensional (E)YMH theory \cite{Julia:1975ff}, \cite{Brihaye:1998cm}.}.

The relations (\ref{rels}), (\ref{rels1}) imply also that for a
vanishing $\gamma$,
the functions $u(r)$ and $J(r)$ vanishes identically and we are lead
to the purely magnetic configurations studied in \cite{Volkov:2001tb}.
\section{Numerical results}
The equations of motion (\ref{e4}) have been solved for a large set of
physical parameters $\alpha$, $\gamma$ and $Q_e$.
The values of $\alpha$ and $Q_e$ are generically sufficient to specify
the system. The various functions, including, e.g.
the value of the parameter $\gamma$, can then be reconstructed from
the numerical solutions.
As expected, the globally regular dyonic solutions have many features in
common with the purely magnetic vortex solutions
discussed in \cite{Volkov:2001tb}; they also present new features that
we will  pointed out in the discussion.
Dyons solutions are found for any monopole
configuration by slowly increasing the
parameter $J(0)$, and extending up
to the maximal value of $\gamma$.

The complete classification of the solutions in the
space of physical parameters $Q_e$, $\alpha$
is a considerable task which is not aimed in this paper.
Instead, we analyzed
in details a few particular classes of solutions which, hopefully,
reflect all the properties of the general pattern.

\newpage
\setlength{\unitlength}{1cm}
\begin{picture}(18,8)
\centering
\put(1,0){\epsfig{file=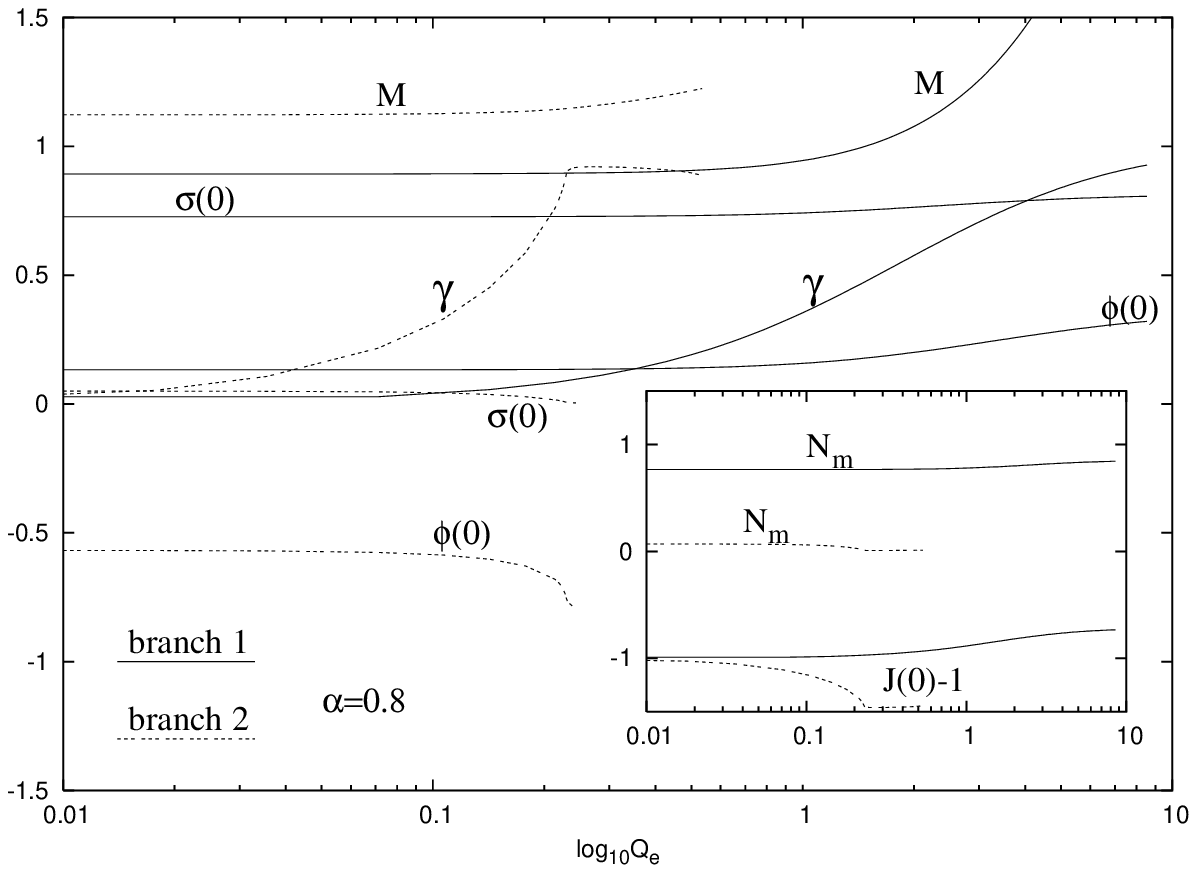,width=14cm}}
\end{picture}
\\
\\
{\small {\bf Figure 1.}
The values  of the parameters $M,~\sigma(0),~N_m,\phi(0),~J(0)$ and
$\gamma$ are shown as a function of
$Q_e$ for two  different branches of solutions at $\alpha=0.8$.}
\subsection{Known solutions}
Let us first briefly summarize the pattern of solution in the known cases
$i.e.$ in the case $\alpha = 0$ and for
a vanishing $A_t$-field, when $\gamma =J(r)=0$.

For $\alpha=0$, the gravity decouples
and we find pure Einstein gravity in $d=5$.
The solutions for the metric ansatz (\ref{metrica}), (\ref{metrica4})
are found
by uplifting a family of four dimensional Einstein-Maxwell-dilaton
black hole solutions discussed in \cite{Garfinkle:1990qj,Gibbons:1987ps}.

However, of interest here is the flat space trivial background ($\phi=J=0$),
in which case we find for the matter part
the BPS  dyon solutions \cite{Prasad:1975kr} uplifted to $d=5$.
Increasing $\alpha$,  we expect these solutions to get deformed by gravity,
at the same time the metric functions $N,~\sigma,~J$
and the dilaton function $\phi$
becoming nontrivial.
The case with gravity present and a vanishing
electric part of the SU(2) field
is studied in  \cite{Volkov:2001tb} (here $J(r)=0$).
The main branch of solutions
exists on the interval $\alpha \in [0, \alpha_{max} ]$ with
$\alpha_{max}\approx 1.268$. Then another branch of solutions
(with higher ADM mass) exist for $\alpha \in [\alpha_{cr}, \alpha_{max}]$
 with
$\alpha_{cr}\approx 0.535$.
Several branches of solutions exist in addition for very small
intervals of $\alpha$ in the region of $\alpha = \alpha_{cr}$.

\subsection{Varying $Q_e$}

Our numerical results reveals that the branches of
solutions occuring in the $A_t=0$ limit are naturally
continued in the dyon case and that for $Q_e \neq 0$ they still exist
for compact intervals of $\alpha$.
A common feature of the dyon solutions is that the parameter $\gamma$
is a monotonically increasing function of the electric charge $Q_e$.

In this subsection
we present the results obtained for fixed $\alpha$ and increasing
the electric charge parameter $Q_e$. 
The case $Q_e$ fixed and $\alpha$
varying will be adressed in the next subsection.
For definiteness we choose $\alpha = 0.8$.
For this value of there
are just two solutions in the purely magnetic case $A_t=0$.
The solution with the lowest ADM mass ($M_{ADM} \approx 0.893$)
will be referred as belonging to the "main branch" while the
solution with the highest ADM mass ($M_{ADM} \approx 1.122$)
will be refered as on the "secondary branch".

In Figure 1, the quantities
$N_{min},~\sigma(0),~J(0),~\phi(0),$ and $u(\infty)\equiv \gamma$
are plotted as functions of $Q_e$
for the main and secondary branches respectively by the
solid and  dotted lines (where $N_{min}$ corresponds to the minimal value of
the metric function $N(r)$).

\newpage
\setlength{\unitlength}{1cm}

\begin{picture}(18,7)
\centering
\put(2,0.0){\epsfig{file=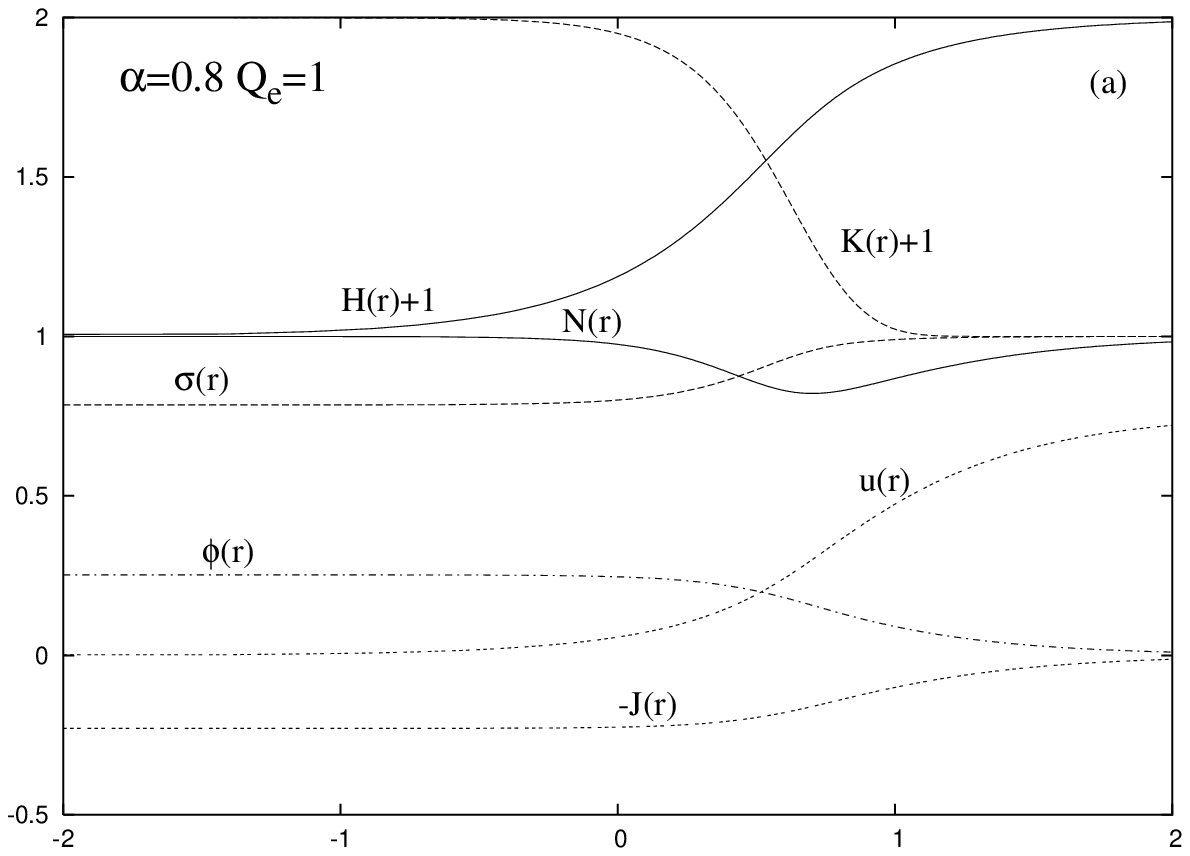,width=11cm}}
\end{picture}
\begin{picture}(19,8.)
\centering
\put(2.6,0.0){\epsfig{file=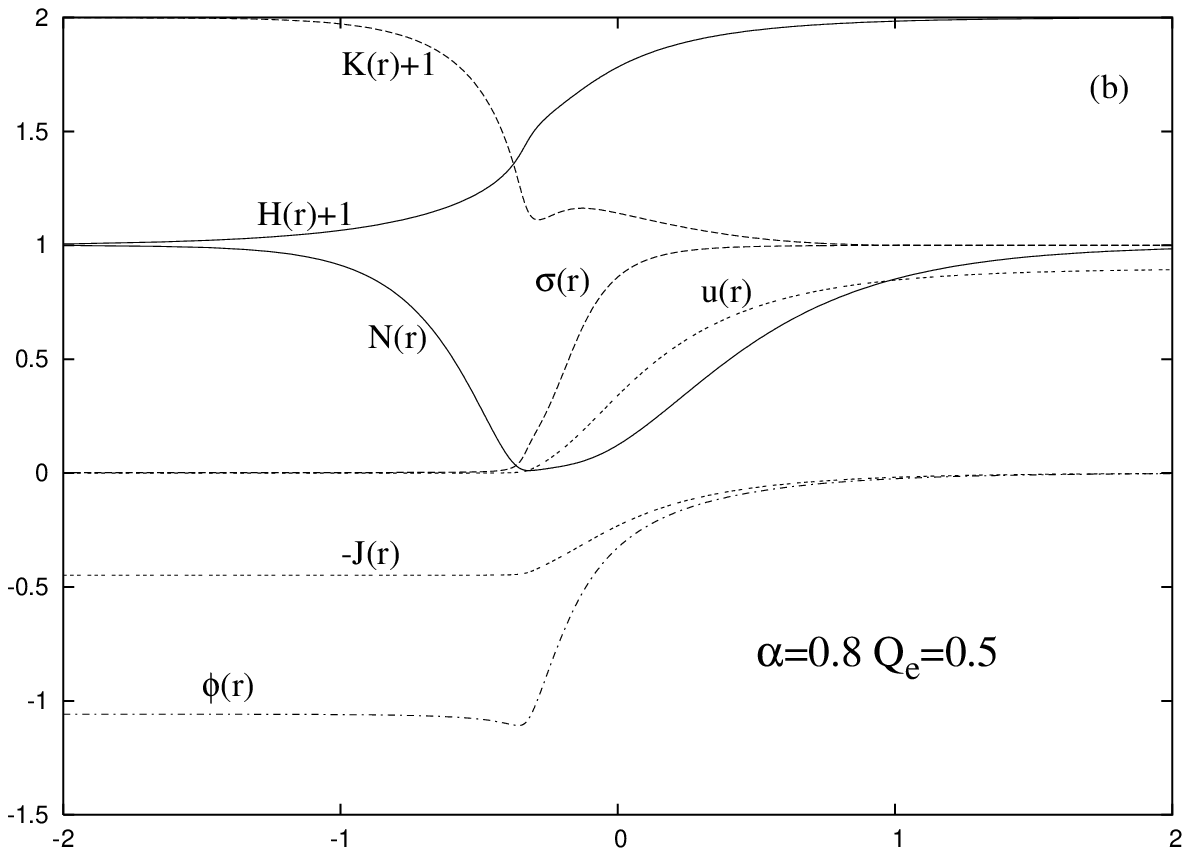,width=11cm}}
\end{picture}
\\
\\
{\small {\bf Figure 2.} The metric functions
$N(r),~\sigma(r),~\phi(r),~J(r)$ and the
matter functions $K(r),H(r)$ and $u(r)$ are shown for $\alpha = 0.1$ and
two values of $Q_e$.}
\\
\\
The ADM mass is also reported by the
line labelled $M$.
It turns out that the main
solution can be constructed without problem
for high values of $Q_e$ (we constructed it up to $Q_e = 10.0)$.

The quantities $M,~N_{min},~\sigma(0),~J(0),~\phi(0)$ all increases
with $Q_e$.
Interestingly, the value $\gamma$ also increases and approaches
 the critical value $\gamma = 1$ for $Q_e \gg 1$.
So our numerical analysis suggests that solutions with
arbitrary
electric charge exist,  composing the main branch.

The construction of the secondary branch of solutions
is, by contrast, more involved. As demonstrated by Figure 1
(dotted lines), the solution
present different behaviours according to the magnitude
of the electric
charge.
For $\alpha= 0.8$, the limit between these two regimes
occurs for $Q_e \approx 0.23$. Let us first analyze the case $Q_e < 0.23$.
For such values of the electric charge,
the parameter $\gamma$ increases
monotonically with $Q_e$ and
culminates at $\gamma \approx 0.92$
for $Q_e \approx 0.23$
(correspondingly $\sigma(0)\approx 0.004$).
The naive picture
would suggest that $\gamma = 1$ will be reached at some 
critical value of $Q_e$, but the scenario turns out to be different.
Indeed,
increasing $Q_e$ such that $Q_e > 0.23$ leads to a new regime
where, namely, both $\gamma$ and $\sigma(0)$ decreases.
\newpage
\setlength{\unitlength}{1cm}
\begin{picture}(18,8)
\centering
\put(1,0){\epsfig{file=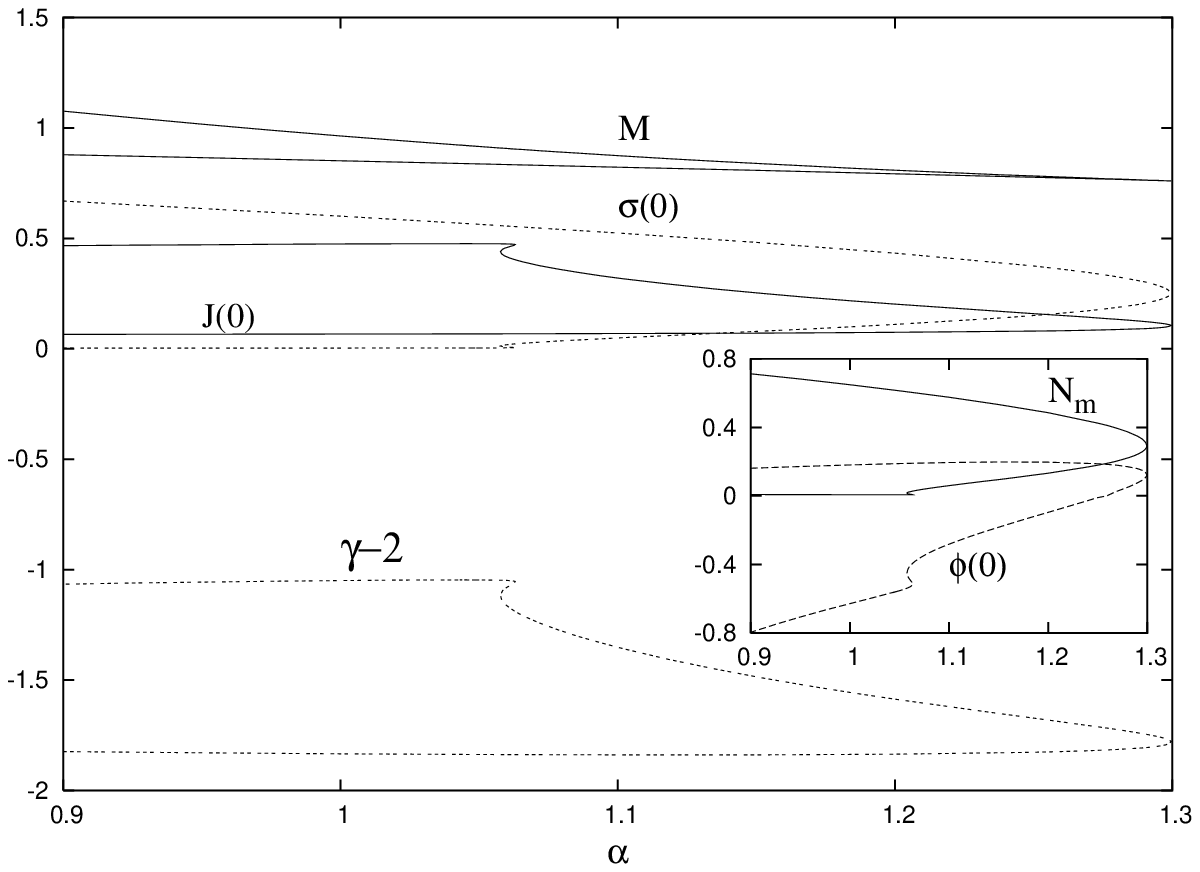,width=14cm}}
\end{picture}
\\
\\
{\small {\bf Figure 3.}
The values  of the parameters $M,~\sigma(0),~N_m,\phi(0),~J(0)$ and
$\gamma$ are shown as a function of
$\alpha$  at $Q_e=0.5$.}
\\
\\

A further numerical study of the equations reveals that the
branch of solutions stop at critical value of $Q_e$
(in the case $\alpha = 0.8$, we find  $Q_{e,c} \approx 0.65$)
 for which the parameter $\sigma(0)$ vanishes.
Here our numerical results suggest that the secondary branch
stops into a singular solution at some maximal value of $Q_e$.

The profiles of the different functions
$N,~\sigma,~K,~H,~\phi,~J,~u$ are
represented in Figure 2 for  generic values
of $Q_E$,
namely $Q=1$ on the main branch  and for $Q_e=0.5$ for the
secondary branch.
On the second graphic, we see in particular that the dilaton
function $\phi(r)$ presents a minimum and that the gauge
function develops an oscillation at $r \sim 0.8$.

\subsection{Varying $\alpha$}

In this section we will pay attention to the
domain of existence of the solutions in
the parameter $\alpha$ for fixed
values of the electric charge.
Our numerical results suggest
 that, for fixed $Q_e$ the solutions exist up to a
maximal value of  $\alpha$, say
$\alpha_{max}$ and that, at least for small values of the electric charge,
this value increases slighly with $Q_e$; for instance~:
\begin{equation}
   \alpha_{max}(Q_e=0) \approx 1.262 \ \ , \ \
       \alpha_{max}(Q_e=0.5) \approx 1.2996 \ \ , \ \
     \alpha_{max}(Q_e=1.0) \approx 1.375 \ \ , \ \
\end{equation}
We studied in details the case corresponding to $Q_e=0.5$,
some relevant data being presented in Figure 3.

We see
in particular that on the main branch
both the minimum of the metric function $N$
and the value of the metric function $\sigma$ at the origin
monotonically decrease when $\alpha$ increases.
The mass parameter $M$ also decreases.
The value of $J(0)$ increases slightly with $\alpha$ on this branch.
For $\alpha > \alpha_{max}$ no solution exist but, along with the case
$Q_e=0$  we found
a second branch of solutions  for $\alpha < \alpha_{max}$.

The solutions on the second branch have a higher ADM mass.
When $\alpha$ increases, the solutions on the second branch
are characterized by increasing values of
the parameters $N_m$ and $\sigma(0)$.
 The parameter $\phi(0)$ become negative at some
intermediate value of $\alpha$. Surprisingly, we observed
on this second branch the occurence of
two supplementary branches of solutions which develop
on a very small interval of the parameter $\alpha$
in the region $\alpha \approx 1.05$. This is
clearly illustrated on Figure 1. The high accuracy of our numerical
routine strongly discard the possibility
of a numerical artefact for  these two extra branches
\footnote{To integrate the equations, we used the differential
equation solver COLSYS which involves a Newton-Raphson method
\cite{COLSYS}.}.

The first integral relations (\ref{rels1})
are also satisfied with a very good accuracy.

Again,
the discussion of the parameter $\gamma$ is crucial and desserves
a particular attention. If we follow the values of this parameter
on the {\it second} branch, it appears that (for increasing $\alpha$)
this parameter
 approaches very closely the value $\gamma = 1$
 (which, remember, turns out to be a critical
point related to the asymptotic behaviour of the function $K(r)$).
However the value $\gamma = 1$ is not attained.
Instead, the very tiny third branch of solution appears there for
$ \alpha \approx 1.06$; from
this value the values of the parameter $\gamma$ decreases and reaches
$\gamma = 0.2217$ at $\alpha = \alpha_{max}$.
To our knowledge, this constitutes a new (at least unexpected)
phenomenon.
It contrasts, namely, with the case of gravitating dyon
studied in \cite{Brihaye:1998cm}
where it was observed that the branch solutions stopped
at some maximal value of the electric charge where the
parameter $\gamma$ reaches the limit $\gamma=1$.

\section{Further remarks}
It is reasonable to expect that the EYM equations
(\ref{einstein-eqs}), ({\ref{YM-eqs}) present also
axially symmetric dyon solutions, obtained within
the general metric ansatz (\ref{metrica})
\footnote{The axially symmetry is defined in
this case with respect to the Killing vector $\partial/\partial \varphi$.}.
These configurations would generalize for  $A_t\neq 0$ the
axially symmetric
monopole vortices discussed in \cite{Brihaye:2002gp}.

An interesting physical question is whether these 
five dimensional EYM vortices can rotate or not.
Working in $d=4$, no nonperturbative
rotating generalizations of the Bartnik-McKinnon solutions
seem to exist \cite{VanderBij:2001nm,Kleihaus:2002ee}.
Here we present arguments that the $d=5$ dyon vortices necessarily have
a vanishing total angular momentum.

The total angular momentum is defined in this case as the charge associated with
the axial Killing vector $\partial/\partial \varphi$, and from (\ref{j5})
we find that, for regular configurations it can be written as a surface
integral at infinity
\begin{eqnarray}
\label{totalJ}
J&=&\int_V T_{\varphi}^{t}\sqrt{-g} d^{3}x
= 2Tr\{\oint_{\infty}dS_{\mu}~WF^{\mu t} \}
\\
\nonumber
&=&2 \pi \lim_{r \rightarrow \infty}  \int_0^{\pi} d \theta \sin \theta^{~}
 r^2[W^{(r)}F^{rt(r)}+W^{(\theta)}F^{rt(\theta)}+W^{(\varphi)}F^{rt(\varphi)}].
\end{eqnarray}
The evaluation of this relation is straightforward.
The boundary conditions at infinity for $A_5$, generalizing (\ref{c2}) 
for the axially symmetric case,
are  $A_5^{(r)}=H_0,~
A_5^{(\theta)}=A_5^{(\varphi)}=0$ \cite{Brihaye:2002gp}.
The requirement of a finite total mass/energy implies that 
the decay of the $F_{rt}^{(a)}$
as $r \to \infty$ is faster than $1/r^{1.5}$.
Also, the contribution
$F_{\varphi 5} ^{(a)} F_{\varphi5}^{(a)} g^{\varphi \varphi}g^{55}$ term
to the total mass/energy is finite if $F_{\varphi 5}^{(a)}$
approaches zero to large $r$ sufficiently rapid.
This implies the fall-off conditions $(W_{\theta}, W_{\varphi})
\sim 1/r^{0.5+\epsilon}$ for large $r$
(with $\epsilon>0$).
Therefore the last two terms in (\ref{totalJ}) give null contribution to the total angular
momentum.
The contribution of the $A_{\varphi}^{(r)}$ to the total angular
momentum is also zero, since it vanishes asymptotically
for a regular configuration.

Therefore we find
\begin{equation}
\label{res-dyon}
 \lim_{r \rightarrow \infty}Tr (r^2 WF_{r t})=-\frac {n  Q_e}{2}  \cos \theta
\end{equation}
(where $Q_e=lim_{r \to \infty}r^2F_{r t}^{(r)}$) and
the total angular momentum of the axially symmetric regular dyon vortices is clearly zero.

This is not a suprise, if we use the observation
that, after the  KK dimensional reduction,
these five dimensional EYM configurations will correspond to $d=4$
EYMHD-U(1) solutions, with a specific coupling between the gauge and Higgs sectors.
However, as found in \cite{VanderBij:2001nm},~\cite{vanderBij:2002sq},
\cite{Volkov:2003ew},
the angular momentum of the $d=4$ pure EYMH  dyons is zero.
The inclusion of a dilaton and a Maxwell field does not change that conclusion.

The $d=5$ solutions can be regarded as the
"regularized" version of the known vacuum black string
configurations, obtained by taking $F_{\mu \nu}=0$ in (\ref{action5}).
Black strings with an SU(2) hair were constructed recently in Ref. \cite{hartmann},
for $A_t=0$ however.
It would be interesting to look for the dyon counterparts
of these configurations,
in particular for axially symmetric solutions.
Different from the regular case, the axially symmetric dyonic black strings
will have a nonvanishing angular momentum localised on the event horizon.
We expect that these configurations (viewed as solutions for the action principle
(\ref{action4})) will share some of the properties of the
pure EYMH solutions discussed recently in \cite{Kleihaus:2004gm}.

For these solutions the geometry of the horizon is $R\times S^2$.
However, in five dimensions more complex configurations
are allowed, as proven by the 
"black ring" solution of the vacuum Einstein equations \cite{Emparan:2001wn}.
This is a rotating black hole with an event horizon of topology 
$S^1\times S^2$, the rotation being required to prevent the ring from collapsing.
 Generalizations with a U(1) gauge field are considered in \cite{Elvang:2003yy}.
Nonabelian versions of these solutions are also likely to exist, 
and will necessarily present an electric part.
In this case one may expect the nonabelian matter content to 
desingularise these solutions, leading to regular,
rotating rings.

\medskip

\noindent
{\bf\large Acknowledgements} \\
ER thanks D. H. Tchrakian for useful discussions.
YB is grateful to the
Belgian FNRS for financial support.
The work of ER is carried out
in the framework of Enterprise--Ireland Basic Science Research Project
SC/2003/390 of Enterprise-Ireland.



%
\end{document}